\documentclass[a4paper,10pt]{article}

\usepackage[latin1]{inputenc}
\usepackage{amsbsy}
\newcommand{\Hil}{{\mathcal H}}
\newcommand{\Ell}{{\mathcal L}}

\newcommand{\Boltz}{ k_{\rm\scriptscriptstyle B}}
\newcommand{\Tr}{{\rm Tr}}
\newcommand{\Ran}{{\rm Ran}}
\newcommand{\Ker}{{\rm Ker}}

\newcommand{\J}{{J\vphantom{\overline{J}}}}
\newcommand{\Jbar}{{\overline{J}}}
\newcommand{\ddt}[1]{{\frac{{\rm d}#1}{{\rm d}t}}}
\newcommand{\DDt}[1]{{\frac{{\rm D}#1}{{\rm D}t}}}
\newcommand{\sq}{{\sqrt{\rho}}}

\newcommand{\sqJ}{{\sqrt{\rho_\J}}}

\newcommand{\sqdotD}{{E_{D\vphantom{\overline{J}}}}}
\newcommand{\sqdotH}{{E_{H\vphantom{\overline{J}}}}}

\newcommand{\sqdot}{{E_{D\vphantom{\overline{J}}}}}
\newcommand{\sqdotJ}{{E_{DJ\vphantom{\overline{J}}}}}

\newcommand{\pardsqdot}[1]{{\frac{{\partial}#1}{{\partial}\sq\sqdot }}}

\newcommand{\cov}[2]{{\langle\Delta #1\Delta #2\rangle}}
\newcommand{\Otimes}{{ \otimes }}
\newcommand{\be}{\begin{equation}}
\newcommand{\ee}{\end{equation}}
\newcommand{\bea}{\begin{eqnarray}}
\newcommand{\eea}{\end{eqnarray}}
\newcommand{\bd}{\begin{displaymath}}
\newcommand{\ed}{\end{displaymath}}
\newcommand{\beastar}{\begin{eqnarray*}}
\newcommand{\eeastar}{\end{eqnarray*}}

\newcommand{\quot}[1]{{``#1''}}
\newcommand{\bsigma}{\mbox{\boldmath $\sigma$}}
\newcommand{\bh}{{\mathbf h}}
\newcommand{\br}{{\mathbf r}}

\title{Positive Nonlinear Dynamical Group Uniting Quantum
Mechanics and Thermodynamics}%
\author{Gian Paolo Beretta}%

\begin{document}
\maketitle
\begin{abstract}
We discuss and motivate the form of the generator of a nonlinear
quantum dynamical group \quot{designed} so as to accomplish a
unification of quantum mechanics (QM) and thermodynamics. We call
this nonrelativistic theory Quantum Thermodynamics (QT). Its
conceptual foundations differ from those of  (von Neumann) quantum
statistical mechanics (QSM) and (Jaynes) quantum information
theory (QIT), but  for  thermodynamic equilibrium (TE) states it
reduces to the same mathematics, and  for zero entropy states it
reduces to  standard unitary  QM. By restricting the discussion to
a strictly isolated system (non-interacting, disentangled and
uncorrelated) we show how the theory departs from the conventional
QSM/QIT rationalization of the second law of thermodynamics. The
nonlinear dynamical group of QT is construed so that the second
law emerges as a theorem of existence and uniqueness of a stable
equilibrium state for each set of mean values of the energy and
the number of constituents. To achieve this QT assumes
$-\Boltz\Tr\rho\ln\rho$ for the physical entropy and is designed
to implement  two fundamental ansatzs. The first is that in
addition to the standard QM states described by idempotent density
operators (zero entropy), a strictly isolated system admits also
states that must be described by non-idempotent density operators
(nonzero entropy). The second is that for such additional states
the law of causal evolution is determined by the simultaneous
action of a Schr\"odinger-von Neumann-type Hamiltonian generator
and a nonlinear dissipative generator which conserves the mean
values of the energy and the number of constituents, and (in
forward time) drives the density operator in the 'direction' of
steepest entropy ascent (maximal entropy increase). The resulting
positive nonlinear dynamical group (not just a semi-group) is
well-defined for all nonequilibrium states, no matter how far from
TE. Existence and uniqueness of solutions of the (Cauchy) initial
state problem for all density operators, implies that the equation
of motion can be solved not only in forward time, to describe
relaxation towards TE, but also backwards in time, to reconstruct
the 'ancestral' or primordial lowest entropy state or limit cycle
from which the system originates.
\end{abstract}

\section{Introduction}

Several authors have attempted to construct a microscopic theory
that includes a formulation of the second law of thermodynamics
\cite{Gorini1,Gorini2,Lindblad,Davies,Spohn,Alicki,Misra,Courbage}.
Some approaches strive to derive irreversibility from a change of
representation of reversible unitary evolution, others from a
change from the von Neumann  entropy functional to other
functionals, or from the loss of information in the transition
from a deterministic system to a probabilistic process, or from
the effect of coupling with one or more heat baths.

We discuss the key elements and features of a different
non-standard theory which introduces {\it de facto} an ansatz of
\quot{intrinsic entropy and instrinsic irreversibility} at the
fundamental level \cite{Frontiers,NATO}, and an additional ansatz
of \quot{steepest entropy ascent} which entails an explicit
well-behaved dynamical principle and the second law of
thermodynamics. To present it, we first discuss an essential
fundamental concept.

\section{States of a strictly isolated individual system}

Let us consider a system $A$ and denote by $R$ the rest of the
universe, so that the Hilbert space of the universe is $ \Hil_{AR}
= \Hil_A\Otimes \Hil_R $. We restrict our attention to a
\quot{strictly isolated} system $A$, by which we mean that at all
times, $-\infty<t<\infty$, $A$ is  uncorrelated (and hence
disentangled) from $R$, i.e., $\rho_{AR}=\rho_{A}\Otimes
\rho_{R}$, and non-interacting, i.e.,  $H_{AR}=H_{A}\Otimes
I_{R}+I_{A}\Otimes H_{R}$.

Many would object at this point that with this premise the following
discussion should be dismissed as useless and unnecessary, because no
``real" system is ever strictly isolated. We reject this argument as
counterproductive, misleading and irrelevant, for we recall that
Physics is a conceptual edifice by which we attempt to model and
unify our perceptions of the empirical world (physical reality
\cite{Margenau}). Abstract concepts such as that of a strictly
isolated system and that of a  state of an individual system not only
are well-defined and conceivable, but have been keystones of
scientific thinking,  indispensable for example to structure the
principle of causality. In what other framework could we introduce,
say, the time-dependent Schr\"odinger equation?

Because the dominant theme of quantum theory is the necessity to
accept that the notion of state involves probabilistic concepts in an
essential way \cite{Park1},  established practices of experimental
science impose that the construct \quot{probability} be linked to the
relative frequency in an \quot{ensemble}. Thus, the purpose of a
quantum theory is to regularize purely probabilistic information
about the measurement results from a \quot{real ensemble} of
identically prepared identical systems. An important scheme for the
classification of ensembles, especially emphasized by von Neumann
\cite{Neumann}, hinges upon the concept of ensemble
\quot{homogeneity}. Given an ensemble it is always possible to
conceive of it as subdivided into many sub-ensembles. An ensemble is
homogeneous iff every conceivable subdivision results into
sub-ensembles all identical to the original (two sub-ensembles are
identical iff upon measurement  on both of the same physical
observable at the same time instant, the  outcomes yield the same
arithmetic mean, and this holds for all conceivable physical
observables). It follows that each individual member system of a
homogeneous ensemble has exactly the same intrinsic characteristics
as any other member, which therefore define the \quot{state} of the
individual system. In other words, the empirical correspondent of the
abstract concept of \quot{state of an individual system} is the
homogeneous ensemble (sometimes also called \quot{pure}
\cite{Bub,Park2,Band} or \quot{proper} \cite{Espagnat,Timpson}).

We restrict our attention to the states of a strictly isolated
individual system. By this we rule out from our present discussion
all heterogeneous preparations, such as those considered in QSM
and QIT, which are obtained by statistical composition of
different homogeneous component preparations. Therefore, we
concentrate on the intrinsic characteristics of each individual
system and their irreducible, non-statistical probabilistic
nature.

\section{Broader quantum kinematics ansatz}

According to standard QM the  states of a strictly isolated
individual system are in one-to-one correspondence with the
one-dimensional orthogonal projection operators on the   Hilbert
space of the system.  We denote such projectors by the symbol $P$. If
$|\psi\rangle$ is an eigenvector of $P$ such that $P|\psi\rangle =
|\psi\rangle$ and $\langle\psi|\psi\rangle = 1$ then $P =
|\psi\rangle\langle\psi|$. It is well known that differently from
classical states, quantum states are characterized by irreducible
intrinsic probabilities. We need not elaborate further on this point.
We only  recall that $-\Tr P\ln P=0$.

Instead, we  adhere to the ansatz \cite{HG} that the set of states in
which a strictly isolated individual system may be found is broader
than conceived in QM, specifically that it is in one-to-one
correspondence with the set of linear operators $\rho$ on $\Hil$,
with $\rho^\dagger=\rho$, $\rho
> 0$, $\Tr\rho = 1$, without the restriction $\rho^2 = \rho$.
We call these the ``state operators" to emphasize that they play
the same role that in QM is played by the projectors $P$, and that
they are associated with the homogeneous preparation schemes. This
fundamental ansatz has been first proposed by Hatsopoulos and
Gyftopoulos \cite{HG}. It allows an implementation of the second
law of thermodynamics at the fundamental level in which the
physical entropy, given by  $s(\rho)=-\Boltz\Tr \rho\ln \rho$,
emerges as an intrinsic microscopic and non-statistical property
of matter, in the same sense as the (mean) energy $e(\rho)=\Tr\rho
H$ is an intrinsic property.

We first assume that our isolated system is an indivisible
constituent of matter, i.e., one of the following:
\begin{itemize}\item A single strictly isolated $d$-level
particle, in which case $\Hil=\Hil_d=\oplus_{k=0}^{d} \Hil_{e_k}$
where $e_k$ is the $k$-th eigenvalue of the (one-particle)
Hamiltonian $H_1$ and $\Hil_{e_k}$ the corresponding eigenspace).
Even if the system is isolated, we do not rule out fluctuations in
energy measurement results and hence we do not assume a
\quot{microcanonical} Hamiltonian (i.e., $H=e_{\tilde
k}P_{\Hil_{e_{\tilde k}}}$ for some ${\tilde k}$) but we assume a
full \quot{canonical} Hamiltonian $H=H_1=\sum_k e_kP_{\Hil_{e_k}}$.
\item A strictly isolated ideal Boltzmann gas of non-interacting
identical indistinguishable $d$-level particles, in which case $\Hil$
is a Fock space, $\Hil={\cal F}_d= \oplus_{n=0}^\infty
\Hil_d^{\otimes n}$. Again, we do not rule out fluctuations in energy
nor in the number of particles, and hence we do not assume a
canonical number operator (i.e., $N={\tilde z}P_{\Hil_d^{\otimes
{\tilde z}}}$ for some ${\tilde z}$) but we assume a full grand
canonical number operator $N=\sum_{n=0}^\infty nP_{\Hil_d^{\otimes
n}}$ and a full Hamiltonian $H=\sum_{n=0}^\infty
H_nP_{\Hil_d^{\otimes n}}$ where $H_n=\sum_{J=1}^n (H_1)_J\otimes
I_{\overline{J}}$ is the $n$-particle Hamiltonian on $\Hil_d^{\otimes
n}$, $(H_1)_J$ denotes the one-particle Hamiltonian on the  $J$-th
particle
 space $(\Hil_d)_J$ and $I_{\overline{J}}$  the identity
operator on the direct product space $\Otimes_{K=1,K\ne
J}^n(\Hil_d)_K$ of all other particles. Note that $[H,N]=0$. \item
A strictly isolated ideal Fermi-Dirac or Bose-Einstein gas of
non-interacting identical indistinguishable $d$-level particles,
in which case $\Hil$ is the antisymmetric or symmetric subspace,
respectively, of the Boltzmann Fock space just defined.
\end{itemize}

We further fix ideas by considering the simplest quantum system, a
2-level particle, a qubit. It is well known \cite{TwoLevel} that
using the 3-vector $\bsigma =(\sigma_1,\sigma_2,\sigma_3)$ of Pauli
spin operators, $[\sigma_j,\sigma_k]=\epsilon_{jk\ell}\sigma_\ell$,
we can represent the Hamiltonian operator as
$H=\hbar\omega({\scriptstyle\frac{1}{2}}I+ \bh \cdot \bsigma)$ where
$\bh$ is a unit-norm 3-vector of real scalars $(h_1,h_2,h_3)$, and
the density operators as
$\rho={\scriptstyle\frac{1}{2}}I+\br\cdot\bsigma$ where $\br$ is a
3-vector of real scalars $(r_1,r_2,r_3)$ with norm $r=|\br|\le 1$,
and $r=1$ iff $\rho$ is idempotent, $\rho^2=\rho$.

If the 2-level particle is strictly isolated, its states in standard
QM are one-to-one with the unit-norm vectors $\psi$ in $\Hil$ or,
equivalently, the unit-trace one-dimensional projection operators on
$\Hil$, $P_\psi=|\psi\rangle\langle\psi |\psi\rangle ^{-1}\langle\psi
|$, i.e., the idempotent density operators $\rho^2=\rho$. Hence, in
the 3-dimensional euclidean space $(r_1,r_2,r_3)$, states map
one-to-one with  points on the unit radius 2-dimensional spherical
surface, $r=1$, the \quot{Bloch sphere}. The mean value of the energy
is $e(\rho)=\Tr \rho H={\scriptstyle\frac{1}{2}}(1+ \bh\cdot\br)$ and
is clearly bounded by $0 \le e(\rho)\le
 \hbar\omega $.  The set of states that share a given mean value
 of the energy are represented  by the
 1-dimensional circular intersection between  the Bloch sphere and
 the constant mean energy plane orthogonal to $\bh$ defined by the $\bh\cdot\br=$
 const condition.
The time evolution according to the Schr\"odinger equation
$\dot\psi=-iH\psi/\hbar$ or, equivalently, $\dot
P_\psi=-i[H,P_\psi]/\hbar$ or \cite{TwoLevel}
$\dot\br=\omega\bh\times\br$ yields a periodic precession of  $\br$
around $\bh$ along such 1-dimensional circular path on the surface of
the Bloch sphere. At the end of every (Poincaré) cycle the strictly
isolated system passes again through its initial state: a clear
pictorial manifestation of the reversibility of Hamiltonian dynamics.

At the level of a strictly isolated qubit, the
Hatsopoulos-Gyftopoulos ansatz amounts to accepting that  the
two-level system admits also states that must be described by points
inside the Bloch sphere, not just  on its surface, even if the qubit
is noninteracting and uncorrelated. The eigenvalues of $\rho$ are
$(1\pm r)/2$, therefore the isoentropic surfaces are concentric
spheres,
\begin{equation} s(\rho)=s(r)=-\Boltz\left( \frac{1+r}{2}\ln
\frac{1+r}{2}+\frac{1-r}{2}\ln\frac{1-r}{2}\right)\ .\end{equation}
The highest entropy state with given mean energy is at the center of
the disk obtained by intersecting the Bloch sphere with the
corresponding constant energy plane. Such states all lie  on the
diameter along the direction of the Hamiltonian vector $\bh$ and are
 thermodynamic equilibrium  (maximum entropy principle
\cite{Book}).

Next, we construct our extension of the Schr\"odinger equation of
motion valid inside the Bloch sphere. By assuming such law of causal
evolution, the second law will emerge as a theorem of the dynamics.

 \section{Steepest-entropy-ascent ansatz}

Let us return to the general formalism for a strictly isolated
system. We go back to the qubit example at the end of the section.

As a first step to force positivity and hermiticity of the state
operator $\rho$ we assume an equation of motion of the form \noindent
\begin{equation}
\label{rhodot}\ddt{\rho}=\rho\,E(\rho)+ E^\dagger(\rho)\,\rho
=\sq\,\Big(\sq E(\rho)\Big)+ \Big(\sq E(\rho)\Big)^\dagger\sq\ ,
\end{equation}
where $E(\rho)$ is a (non-hermitian) operator-valued (nonlinear)
function of $\rho$ that we call
 the
\quot{evolution} operator. Without loss of generality, we write
$E=E_+ +iE_-$ where $E_+=(E+E^\dagger)/2$ and $E_-=(E-E^\dagger)/2i$
are hermitian operators, so that Eq.\ (\ref{rhodot}) takes the form
\begin{equation}
\label{rhodot2}\ddt{\rho}=-i[E_-(\rho),\rho]+ \{E_+(\rho),\rho\} \
,
\end{equation}with $[\,\cdot\,,\,\cdot\,]$ and  $\{\,\cdot\,,\,\cdot\,\}$
the usual commutator and anti-commutator, respectively.

We consider the space of linear (not necessarily hermitian) operators
on $\Hil$ equipped with the real scalar product
\begin{equation}
\label{Scalar} (F|G) = \Tr (F^\dagger G + G^\dagger F)/2 \ ,
\end{equation}
so that for any time-independent hermitian observable $R$ on $\Hil$,
the rate of change of the mean value  $r(\rho)=\Tr (\rho R)=(\sq|\sq
R)$ can be written as
\begin{equation}
\label{rateR} \ddt{r(\rho)}=\Tr( \ddt{\rho}R)=2\left.\left(\sq
E\right|\sq R\right) \ ,
\end{equation}
from which it follows  that a set of $r_i(\rho)$'s is time
invariant iff $\sq E$ is orthogonal to the linear span of the set
of operators $\sq R_i$, that we denote by $\Ell {\{\sq R_{i }
\}}$.

For an isolated system, we therefore require that, for every
$\rho$, operator $\sq E$ be orthogonal [in the sense of scalar
product (\ref{Scalar})] to the linear manifold $\Ell {\{\sq R_{i }
\}}$ where the set $\{\sq R_{i } \}$ always includes
 $\sq I$, to preserve $\Tr \rho=1$, and
$\sq H$, to conserve the mean energy $e(\rho)=\Tr \rho H$. For a
field of indistinguishable particles we  also include $\sq N$ to
conserve the mean number of particles $n(\rho)=\Tr \rho N$. For a
free particle we would include $\sq P_x$, $\sq P_y$, $\sq P_z$ to
conserve the mean momentum vector ${\mathbf p}(\rho)=\Tr \rho
{\mathbf P}$, but here we omit this case for simplicity
\cite{Gheorghiu}.

Similarly, the rate of change of the entropy  functional can be
written as
\begin{equation}
\label{rateS} \ddt{s(\rho)}=  \left(\sq E\left|-2\Boltz\left[\sq
+\sq\ln\rho\right]\right.\right) \ ,
\end{equation}
where the operator $-2\Boltz\left[\sq +\sq\ln\rho\right]$ may be
interpreted as the gradient  (in the sense of the functional
derivative) of the entropy functional $s(\rho)=-\Boltz\Tr\rho\ln\rho$
with respect to  operator $\sq$ (for the reasons why in our theory
the physical entropy is represented by the von Neumann functional,
see Refs. \cite{HG,Cubukcu}).

 It is noteworthy that the Hamiltonian evolution operator
\begin{equation}\label{sqHamil}
\sqdotH=i H/\hbar \ ,
\end{equation}
is such that $\sq\sqdotH$ is orthogonal to $\Ell {\{\sq I,\sq H (,
\sq N) \}}$ as well as to the entropy gradient operator
$-2\Boltz\left[\sq +\sq\ln\rho\right]$. It yields a
Schr\"odinger-Liouville-von Neumann unitary dynamics
\begin{equation}
\label{rhodotH}\ddt{\rho}=\rho\,E_H+ E_H^\dagger\rho
=-\frac{i}{\hbar}[H,\rho] \ ,
\end{equation}
which maintains  time-invariant all the eigenvalues of $\rho$.
Because of this feature,  all time-invariant (equilibrium) density
operators according to Eq. (\ref{rhodotH})  (those that commute with
$H$) are globally stable \cite{Lyapunov} with respect to
perturbations that do not alter the mean energy (and the mean number
of particles). As a result, for given values of the mean energy
$e(\rho)$ and the mean number of particles $n(\rho)$ such a dynamics
would in general imply many stable equilibrium states, contrary to
the second law requirement that there must be only one (this is the
well-known Hatsopoulos-Keenan statement of the second law \cite{HK},
which entails  \cite{Book} the other well-known statements by
Clausius, Kelvin, and Carath\'eodory).

Therefore, we assume that in addition to the Hamiltonian term
$\sqdotH$, the evolution operator $E$ has an additional component
$\sqdotD$,
\begin{equation}
\label{sqdot}E =\sqdotH + \sqdotD \ ,
\end{equation}
that we will take so that $\sq\sqdotD$ is at any $\rho$ orthogonal
both to $\sq\sqdotH$  and to the intersection of the
 linear manifold $\Ell {\{\sq R_{i }
\}}$  with the isoentropic hypersurface to which $\rho$ belongs
(for a two level system, such intersection is  a one-dimensional
planar circle inside the Bloch sphere). In other words, we assume
that $\sq\sqdotD$ is proportional to the component of the entropy
gradient operator $-2\Boltz\left[\sq +\sq\ln\rho\right]$
orthogonal to $\Ell {\{\sq R_{i } \}}$,
\begin{equation} \sq\sqdotD
=-\frac{1}{2 \tau(\rho)}\left[\sq\ln\rho \right]_{\bot \Ell {\{\sq
I,\sq H (, \sq N) \}}} \ ,
\end{equation}
where we  denote the \quot{constant} of proportionality by
$1/2\tau(\rho)$ and use the fact that $\sq$ has no component
orthogonal to $\Ell {\{\sq I,\sq H (, \sq N) \}}$.

It is important to note that the \quot{intrinsic dissipation} or
\quot{intrinsic relaxation} characteristic time $\tau(\rho)$ is left
unspecified in our construction and need not be a constant. All our
results  hold as well if  $\tau(\rho)$ is some reasonably well
behaved positive definite functional of
 $\rho$. The empirical and/or theoretical determination of $\tau(\rho)$ is
 a most challenging open problem in our research program.  For
 example, it has been suggested \cite{CubukcuThesis} that the
experiments by Franzen \cite{Franzen} (intended to evaluate the
spin relaxation time constant of vapor under vanishing pressure
conditions)  and by Kukolich \cite{Kukolich} (intended to provide
a laboratory validation of the time-dependent Schr\"odinger
equation) both suggest some evidence of an intrinsic relaxation
time.

Using standard geometrical notions, we can show
\cite{Frontiers,NATO,thesis,ArXiv1} that given any set of linearly
independent  operators  $\{\sq R_{i } \}$ spanning $\Ell {\{\sq I,\sq
H (, \sq N) \}}$  the dissipative evolution operator takes the
explicit expression
\begin{equation}\label{ED} \sq\sqdotD=\frac{1}{2\Boltz
\tau(\rho)\,}\sq\Delta M(\rho) \end{equation} where $M(\rho)$ is a
\quot{Massieu-function} operator defined by
\begin{equation}\label{Massieu}M(\rho)=\frac{\left|
\begin{array}{ccccc}  S &  R_1 &\!\cdots\!&
 R_i &\!\cdots\!\\ \\ {\scriptstyle\cov{S}{R_1}}&
{\scriptstyle\cov{R_1}{R_1}} &\!\cdots\!&
{\scriptstyle\cov{R_i}{R_1}} & \!\cdots\!\\ \vdots & \vdots &
\!\ddots\! & \vdots&\!\ddots\!\\ {\scriptstyle\cov{S}{R_i}}&
{\scriptstyle\cov{R_1}{R_i}} &\!\cdots\!&
{\scriptstyle\cov{R_i}{R_i}} & \!\cdots\!\\ \vdots& \vdots &
\!\ddots\! & \vdots&\!\ddots\! \end{array} \right|} {\Gamma(\{\sq
R_{i}\}) }  \ ,
\end{equation}
and  we use the notation ($F$ and $G$ hermitian)
\begin{eqnarray}
S&=&-\Boltz P_{\Ran\, \rho} \ln\rho \ ,\\ \label{Delta}\Delta F&=&F-
\Tr(\rho F)I \ ,\\ \label{Cov}\cov{F}{G}&=&  (\sq \Delta F|\sq \Delta
G) =\frac{1}{2} \Tr(\rho\{\Delta F,\Delta G\})\ ,\\ \Gamma(\{\sq
R_i\})&=&\det[\cov{R_i}{R_j}] \quad \mbox{(a Gram determinant)}\ .
\end{eqnarray}
The Massieu-function operator defined by Eq. (\ref{Massieu})
generalizes to any non-equilibrium state the well-known equilibrium
Massieu characteristic function $s(\rho_{TE})-\beta\, e(\rho_{TE})\,
[+\beta\mu\, n(\rho_{TE})]$. As a result, our full equation of motion
takes the form
\begin{equation}
\label{rhodotHM}\ddt{\rho}=-\frac{i}{\hbar}[H,\rho]+
\frac{1}{2\Boltz\tau(\rho)}\{\Delta M(\rho),\rho\} \ .
\end{equation}

Gheorghiu-Svirschevski \cite{Gheorghiu} re-derived our
 nonlinear equation of motion from a  variational principle that in our notation  may
 be cast as follows \cite{ArXiv1},
\begin{equation} \max_{\sq\sqdot}\
\ddt{s(\rho)} \mbox{ subject to }\ddt{r_i(\rho)}=0 \ {\rm and\ }
(\sq \sqdot | \sq \sqdot)=c^2(\rho)\ ,
\end{equation}
where $r_0(\rho)=\Tr\rho$, $r_1(\rho)=\Tr H\rho$ [, $r_2(\rho)=\Tr
N\rho$], and $c^2(\rho)$ is
 some  positive functional. The last constraint means that we are not
 really searching for  \quot{maximal entropy production} but only
  for the direction of steepest entropy ascent, leaving
 unspecified the rate at which such direction \quot{attractsd} the state
 of the
 system. The necessary
condition in terms of Lagrange multipliers is \begin{equation}
\pardsqdot{}\ddt{s}-\sum_i \lambda_i
\pardsqdot{}\ddt{r_i}-\lambda_0 \pardsqdot{} (\sq\sqdot |\sq \sqdot) =0
\ , \end{equation} and, using Eqs. (\ref{rateR}) and (\ref{rateS})
becomes
\begin{equation}\label{maxS}-2\Boltz( \sq+ \sq\ln\rho)-2\sum \lambda_i
\sq R_i-2\lambda_0 \sq\sqdot=0 \ , \end{equation} which inserted
in the constraints and solved for the multipliers yields Eq.
(\ref{ED}).

The resulting rate of entropy change (entropy generation by
irreversibility, for the system is isolated) is given by the
equivalent expressions
 \begin{eqnarray}
\ddt{s(\rho)}&=&-\Boltz\ddt{\Tr\rho\ln\rho}=
{4\Boltz\tau(\rho)}\left(\sq\sqdotD \left|\sq\sqdotD
\right.\right)
\\ &=& \frac{\Boltz }{ \tau(\rho)} \frac{\Gamma(\sq\ln\rho,\{\sq
R_i\})}{\Gamma(\{\sq R_i\})}=
 \frac{1 }{ \Boltz\tau(\rho)} \frac{\Gamma(\sq S,\{\sq R_i\})}{\Gamma(\{\sq
R_i\})}\ge 0 \ .
\end{eqnarray}
Because a Gram determinant $\Gamma(\sq X_1,\dots,\sq
X_N)=\det[\cov{X_i}{X_j}] $ is either strictly positive or zero iff
operators $\{\sq X_i\}$ are linearly dependent, the rate of entropy
generation is either a positive semi-definite nonlinear functional of
$\rho$, or it is zero iff operators $\sq S, \sq I,\sq H (, \sq N) $
are linearly dependent, i.e., iff the state operator is of the form
 \begin{equation}\label{nondiss} \rho
= \frac{B \exp[-\beta H\,(+\nu N)]B}{\Tr( B \exp[-\beta H\,(+\nu
N)])}\ ,
\end{equation} for some \quot{binary} projection operator $B$
($B^2=B$, eigenvalues either 0 or 1) and some real scalar(s) $\beta$
(and $\nu$).  Non-dissipative states are therefore all and only the
density operators that have the nonzero eigenvalues
\quot{canonically} (or \quot{grand canonically}) distributed. For
them, $\sq\sqdotD=0$ and our equation of motion (\ref{rhodotHM})
reduces to the Schr\"odinger--von Neumann form
$i\hbar\dot\rho=[H,\rho]$. Such states are either equilibrium states,
if $[B,H]= 0$, or belong to a limit cycle and undergo a unitary
hamiltonian dynamics,   if $[B,H]\ne 0$, in which case
\begin{eqnarray} &\rho(t) = B(t) \exp[-\beta H\,(+\nu N)]B(t)/\Tr[B(t)
\exp[-\beta H\,(+\nu N)]]\ ,&\\ \label{Bunitary}&B(t) = U(t) B(0)
U^{-1}(t) \ ,\quad U(t)=\exp(-itH /\hbar)\ .& \end{eqnarray} For $\Tr
B=1$ the states (\ref{nondiss}) reduce to the (zero entropy) states
of standard QM, and  obey the standard unitary dynamics generated by
the usual time-dependent Schr\"odinger equation. For $B=I$ we have
the maximal-entropy (thermodynamic-equilibrium) states, which turn
out to be the only globally stable equilibrium states of our
dynamics, so that the Hatsopoulos-Keenan statement of the second law
emerges as an exact and general dynamical theorem.

Indeed, in the framework of our extended theory,  all equilibrium
states and limit cycles that have at least one null eigenvalue of
$\rho $ are unstable. This is because any neighboring state
operator with one of the null eigenvalues perturbed (i.e.,
slightly \quot{populated}) to a small value $\epsilon$ (while some
other eigenvalues are slightly changed so as to ensure that the
perturbation preserves the mean energy and the mean number of
constituents), would eventually proceed \quot{far away} towards a
new partially-maximal-entropy state or limit cycle with a
canonical distribution which fully involves also the newly
\quot{populated} eigenvalue while the other null eigenvalues
remain zero.

It is clear that the canonical (grand-canonical) density operators
$\rho_{TE} =  \exp[-\beta H\,(+\nu N)]/\Tr(  \exp[-\beta H\,(+\nu
N)])$ are the only stable equilibrium states, i.e., the TE states of
the strictly isolated system. They are mathematically identical to
the density operators which also in QSM and QIT are  associated with
TE, on the basis of their maximizing the von Neumann indicator of
statistical uncertainty $-\Tr\rho\ln\rho$ subject to given values of
$\Tr H\rho$ (and $\Tr N\rho$). Because maximal entropy mathematics in
QSM and QIT  successfully represents TE physical reality,  our
theory, by entailing the same mathematics for the stable equilibrium
states, preserves all the successful results of equilibrium QSM and
QIT. However, within QT such  mathematics takes up an entirely
different physical meaning. Indeed,  each density operator here does
not represent statistics of measurement results from a
\quot{heterogeneous} ensemble, as in QSM and QIT where, according to
von Neumann's recipe \cite{Neumann,MPLA2}, the \quot{intrinsic}
uncertainties (irreducibly introduced by standard QM) are mixed with
the \quot{extrinsic} uncertainties  (related to the heterogeneity of
its preparation, i.e., to not knowing the exact state of each
individual system in the ensemble). In QT, instead, each density
operator, including the maximal-entropy stable TE ones, represents
\quot{intrinsic} uncertainties only,  because it is associated with a
homogeneous preparation and, therefore, it represents the state of
each and every individual system of the homogeneous ensemble.

 We noted elsewhere \cite{MPLA} that the
fact that our nonlinear equation of motion preserves the null
eigenvalues of $\rho$, i.e., conserves the cardinality $
\dim\Ker(\rho)$ of the set of zero eigenvalues, is an important
physical feature consistent with recent experimental tests (see the
discussion of this point in Ref. \cite{Gheorghiu} and references
therein) that rule out, for pure (zero entropy) states, deviations
from linear and unitary dynamics and confirm that initially
unoccupied eigenstates cannot spontaneously become occupied. This
fact, however, adds nontrivial experimental and conceptual
difficulties to the problem of designing fundamental tests capable,
for example, of ascertaining whether decoherence originates from
uncontrolled interactions with the environment due to the practical
impossibility of obtaining strict isolation, or else it is a more
fundamental intrinsic feature of microscopic dynamics requiring an
extension of QM like the one we propose.

For a confined, strictly isolated  $d$-level system, our equation of
motion for non-zero entropy states ($\rho^2\ne\rho$) takes the
following forms \cite{TwoLevel,PRE}. If the Hamiltonian is fully
degenerate [$H=eI$, $e(\rho)=e$ for every $\rho$],
\begin{equation}
\ddt\rho =-\frac{i}{\hbar}[H,\rho]-\frac{1}{\tau}
\left(\rho\ln\rho-\rho\,\Tr\rho\ln\rho\right) \ ,
\end{equation}
while if the Hamiltonian is nondegenerate,
\begin{equation}
\ddt\rho =-\frac{i}{\hbar}[H,\rho]-\frac{1}{\tau} \frac { \left|
\begin{array}{ccccc} {\rho\ln\rho } & & {\rho} & & \frac
 {1}{2}
\{ { H},\rho \}
\bigskip \nonumber \\
{\Tr\rho \ln\rho }& & {\displaystyle 1} &  & {\Tr\rho { H} }\
\bigskip \nonumber \\
 { \Tr\rho{ H}\ln\rho } &
&  {\Tr\rho { H} } & & {\Tr\rho { H}^2 } \nonumber \\
\end{array}
\right| } { \Tr\rho { H}^2-(\Tr\rho { H})^2 } \ .\nonumber \\
 \nonumber
\\
\end{equation}
In particular, for a non-degenerate two-level system, it may be
expressed in terms of the Bloch sphere representation (for
$0<r<1$) as \cite{TwoLevel}
\begin{equation}\dot\br=\omega\bh\times\br
-\frac{1}{\tau}\left(\frac{1-r^2}{2r}\ln\frac{1-r}{1+r}\right)
\frac{\bh\times\br\times\bh}{1-(\bh\cdot\br)^2}
\end{equation}
from which it is clear that the dissipative term lies in the
constant mean energy plane and is directed towards the axis of the
Bloch sphere identified by the Hamiltonian vector $\bh$. The
nonlinearity of the equation does not allow a general explicit
solution, but on the central constant-energy plane, i.e., for
initial states with $\br\cdot\bh=0$, the equation implies
\cite{TwoLevel}
\begin{equation}\ddt{}\ln\frac{1-r}{1+r}=-\frac{1}{\tau}\ln\frac{1-r}{1+r}
\end{equation} which, if $\tau$ is constant, has the solution
\begin{equation}r(t)=\tanh\left[-\exp\left(-\frac{t}{\tau}\right)
\ln\frac{1-r(0)}{1+r(0)}\right] \ .
\end{equation}
This, superposed with the  precession around the hamiltonian
vector, results in a spiraling  approach to the maximal entropy
state (with entropy $\Boltz\ln 2$). Notice, that the spiraling
trajectory is well-defined and within the Bloch sphere for all
times $-\infty<t<+\infty$, and if we follow it backwards in time
it approaches as $t\to-\infty$ the limit cycle which represents
the standard QM (zero entropy) states evolving according to the
Schr\"odinger equation.

This example shows quite explicitly a general feature of our
 nonlinear equation of motion which follows from the
existence and uniqueness of its solutions for any initial density
operator both in forward and backward time. This feature is a
consequence of two facts: (1) that zero eigenvalues of $\rho$ remain
zero and therefore no eigenvalue can cross zero and become negative,
and (2) that $\Tr\rho$ is preserved and therefore if initially one it
remains one. Thus, the eigenvalues of $\rho$ remain positive and less
than unity. On the conceptual side,  it is also clear that our theory
implements a strong causality principle by which all future as well
as all past states are fully determined by the present state of the
isolated system, and yet the dynamics is physically
(thermodynamically) irreversible. Said differently, if we formally
represent the general solution of the Cauchy problem by
$\rho(t)=\Lambda_t\rho(0)$ the nonlinear map $\Lambda_t$ is a group,
i.e., $\Lambda_{t+u}=\Lambda_t\Lambda_u$ for all $t$ and $u$,
positive and negative. The map is therefore \quot{invertible}, in the
sense that $\Lambda_{-t}=\Lambda_t^{-1}$, where the inverse map is
defined by $\rho(0)=\Lambda^{-1}_t\rho(t)$.

It is a nontrivial observation that the non-invertibility of the
dynamical map is not at all necessary to represent a physically
irreversible dynamics. Yet, innumerable attempts to build
irreversible theories start from the assertion that in order to
represent thermodynamic irreversibility the dynamical map should be
non-invertible. The arrow of time in our view is not to be sought for
in the impossibility to retrace past history, but in the  spontaneous
tendency of any physical system to internally redistribute its energy
(and, depending on the system, its other conserved properties such
number of particles, momentum, angular momentum) along the path of
steepest entropy ascent.

\section{Onsager reciprocity}

The intrinsically irreversible dynamics entailed by the dissipative
(non-hamiltonian) part of our nonlinear equation of motion also
entails an Onsager reciprocity theorem. To see this, we first note
 that any density operator $\rho$ can be written as
\cite{Onsager}
\begin{equation}\label{anystate} \rho =\frac{ B\exp(-\sum_j f_j
X_j)B}{\Tr B\exp(-\sum_j f_j X_j)} \ , \end{equation} where the
possibly time-dependent Boolean $B$ is such that $B=P_{\Ran{\rho}}$
($=I-P_{\Ker{\rho}}$) and the time-independent operators $X_j$
together with the identity $I$ form a set such that their
restrictions to $\Hil'=B\Hil$, $\{I',X'_j\}$ span the real space  of
hermitian operators on $\Hil'=B\Hil$. Hence,
 \begin{eqnarray}\label{sqlnrho} &\sq\ln\rho=-f_0
\sq-\sum_j f_j \sq X_j  \ ,&\\ & x_j (\rho)=\Tr (\rho X_j)\ ,&\\
&s(\rho)=\Boltz f_0 + \Boltz\sum_j f_j\, x_j (\rho) \label{DDts} \
,&\\ & {\rm where\quad}\displaystyle\Boltz f_j =
\left.\frac{\partial s(\rho)}{\partial x_j(\rho)}\right|_{x_{i\ne
j}(\rho)}\label{DAf}&
\end{eqnarray}  may be interpreted as a \quot{generalized affinity} or
force. Defining
\begin{equation}\label{dissratexi} \DDt{x_i(\rho)}
=2\!\left.\left(\sqdotD\right|\sq X_i\right)\; .
\end{equation}
as \quot{the dissipative rate of change} of the mean value $x_j
(\rho)$, we find
 \begin{equation}\label{dissrate}
\DDt{x_i(\rho)} =\sum_j f_j \, L_{ij}(\rho)\ , \end{equation} where
the coefficients $L_{ij}(\rho)$ (nonlinear in $\rho$) may be
interpreted as \quot{generalized conductivities} and are given
explicitly (no matter how far $\rho$ is  from TE) by
\label{Lij}\begin{eqnarray} L_{i j}(\rho) &=& L_{ji}(\rho)
={\displaystyle \frac{1}{\tau(\rho)}} \!\left.\left([\sq X_i
]_{\bot\Ell {\{\sq R_{i } \}}}\right|[\sq X_j ]_{\bot\Ell {\{\sq R_{i
} \}}}\right)\label{gramL} \\ &=&{\displaystyle \frac{1}{\tau(\rho)}}
\frac{\left|
\begin{array}{ccccc} {\scriptstyle\cov{X_i}{X_j}}
&{\scriptstyle\cov{R_1}{ X_j }} &\!\cdots\!& {\scriptstyle\cov{R_k}{
X_j }} & \!\cdots\!\\ \\ {\scriptstyle\cov{ X_i }{R_1}}
&{\scriptstyle\cov{R_1}{R_1}} &\!\cdots\!&
{\scriptstyle\cov{R_k}{R_1}} & \!\cdots\!\\ \vdots &\vdots &
\!\ddots\! & \vdots&\!\ddots\!\\ {\scriptstyle\cov{ X_i }{R_k}}
&{\scriptstyle\cov{R_1}{R_k}} &\!\cdots\!&
{\scriptstyle\cov{R_k}{R_k}} & \!\cdots\!\\ \vdots &\vdots &
\!\ddots\! & \vdots&\!\ddots\! \end{array} \right| }{\displaystyle
\Gamma(\{\sq R_k\})} \ , \end{eqnarray} and therefore form a
symmetric, non-negative definite Gram matrix $[ L_{i j}(\rho)]$,
which is strictly positive iff all operators $ [\sq X_i ]_{\bot\Ell
{\{\sq R_{i } \}}} $ are linearly independent.

The rate of entropy generation may be rewritten as a quadratic
form of the generalized affinities, \begin{equation}\label{sdotL}
\ddt{s(\rho)} =\Boltz\sum_i\sum_j f_i f_j L_{i j}(\rho)\ .
\end{equation} If all operators $ [\sq X_i
]_{\bot\Ell {\{\sq R_{i } \}}} $ are linearly independent, $\det[
L_{i j}(\rho)]\ne 0$ and Eq.\ (\ref{dissrate}) may be solved to yield
\begin{equation} f_j=\sum_i L^{-1}_{i j}(\rho)\DDt{x_i(\rho)} \ ,
\end{equation} and the rate of entropy generation can be written also as a
quadratic form of the dissipative rates
\begin{equation} \ddt{s(\rho)} =\Boltz\sum_i\sum_j
L^{-1}_{ij}(\rho)\DDt{x_i(\rho)} \DDt{x_j(\rho)}\ . \end{equation}

\section{Composite systems and reduced dynamics}

The composition of the system is embedded in the structure of the
Hilbert space as a direct product of the subspaces associated with
the individual elementary constituent subsystems, as well as in
the form of the Hamiltonian operator. In this section, we consider
a system composed of distinguishable and indivisible elementary
constituent subsystems. For example:
\begin{itemize}
\item A strictly isolated composite of $r$ distinguishable
$d$-level particles, in which case
$\Hil=\otimes_{J=1}^r\Hil_{d_J}$ and $H=\sum_{J=1}^r
(H_1)_J\otimes I_{\overline{J}}+V$ where $V$ is some interaction
operator over $\Hil$.
\item A strictly isolated ideal mixture of $r$ types of Boltzmann,
Fermi-Dirac or Bose-Einstein gases of non-interacting identical
indistinguishable $d_J$-level particles, $J=1,\dots,r$, in which case
$\Hil$ is a composite of Fock spaces $\Hil=\otimes_{J=1}^r{\cal
F}_{d_J}=\oplus_{n_1=0}^\infty
 \cdots \oplus_{n_r=0}^\infty \Hil_{d_1}^{\otimes n_1}
 \otimes \cdots\otimes\Hil_{d_r}^{\otimes n_r}$ where the
factor Fock spaces belonging to  Fermi-Dirac (Bose-Einstein)
components are restricted to their antisymmetric (symmetric)
subspaces. Again, we assume full grand-canonical number operators
$N_J=\sum_{n_J=0}^\infty n_JP_{\Hil_{d_J}^{\otimes n_J}}$ and
Hamiltonian $H=\sum_{J=1}^rI_{\overline{J}}\otimes\sum_{n_J=0}^\infty
H_{n_J}P_{\Hil_{d_J}^{\otimes n_J}}+V$.
\end{itemize}

For compactness of notation we denote the subsystem Hilbert spaces
as
\begin{equation}
\label{HilbertSpace}\Hil  = \Hil^1\Otimes
\Hil^2\Otimes\cdots\Otimes\Hil^r =\Hil^\J\Otimes\Hil^\Jbar\ ,
\end{equation}
where $\Jbar$ denotes all subsystems except the $J$-th one. The
overall system is strictly isolated in the sense already defined,
and the Hamiltonian operator
\begin{equation}
\label{Hamiltonian} H  = \sum_{J=1}^r H_\J\Otimes I_\Jbar + V \ ,
\end{equation} where $H_\J$ is the Hamiltonian  on
$\Hil^\J$ associated with the $J$-th subsystem when isolated and
$V$ (on $\Hil$) the interaction Hamiltonian among the $r$
subsystems.

The subdivision into elementary constituents, considered as
indivisible, and reflected by the structure of the Hilbert space
$\Hil$ as a direct product of subspaces, is particularly important
because it defines the level of description of the system and
specifies its elementary structure.  The system's internal structure
we just defined determines the form of the nonlinear dynamical law
proposed by this author \cite{thesis,ArXiv1,Cimento} to implement the
steepest entropy ascent ansatz in a way compatible with the obvious
self-consistency \quot{separability} and \quot{locality} requirements
\cite{MPLA}. It is important to note that, because our dynamical
principle is nonlinear in the density operator, we cannot expect the
form of the equation of motion to be independent of the system's
internal structure.

The equation of motion that we \quot{designed} in
\cite{thesis,Cimento} so as to guarantee all the necessary
features (that we list in Ref. \cite{MPLA}), is
\begin{equation}
\label{rhodotHMJ}\ddt{\rho}=-\frac{i}{\hbar}[H,\rho]+\sum_{J=1}^r
\frac{1}{2\Boltz\tau_\J(\rho)}\{(\Delta
M_\J(\rho))^\J,\rho_\J\}\Otimes \rho_\Jbar \ ,
\end{equation}
where we use the notation [see Ref. \cite{ArXiv1} for interpretation
of $(S)^\J$ and $(H)^\J$ ]
\begin{eqnarray}
  \sqJ(\Delta
M_\J(\rho))^\J &=& \left[\sqJ(S)^\J \right]_{ \bot\Ell
{\{\sqrt{\rho_\J}(R_{i\J})^J\}}}\ ,\\ \Ell
\{\sqrt{\rho_\J}(R_{i\J})^J\} &=&\mbox{ lin. span of }
\sqrt{\rho_\J}I_\J, \sqrt{\rho_\J}(H_{\J})^J,
\sqrt{\rho_\J}(N_{k\J})^J \\ ( F_\J | G_\J)_\J&=& \Tr_\J(
F^\dagger_\J G_\J +G^\dagger_\J F_\J)/2 \ , \\
(R_{i\J})^\J&=&\Tr_\Jbar [(I_\J\Otimes \rho_\Jbar) R_{i\J}]\ ,
\\ (S)^\J&=&\Tr_\Jbar [(I_\J\Otimes \rho_\Jbar) S]\ ,
\end{eqnarray}
and the \quot{internal redistribution characteristic times}
$\tau_\J(\rho)$'s are some positive constants or positive
functionals of the overall system's density operator $\rho$.

All the results found for the single constituent extend in a
natural way to the composite system. For example, the rate of
entropy change becomes
\begin{equation}
\ddt{s(\rho)}= \sum_{J=1}^r \frac{1 }{ \Boltz\tau_\J(\rho)}
\frac{\Gamma(\sqJ (S)^\J,\{\sqJ (R_{i\J})^\J \})}{\Gamma(\{\sqJ
(R_{i\J})^\J \})} \ .
\end{equation}

The dynamics reduces to the Schr\"odinger-von Neumann unitary
Hamiltonian dynamics when, for each $J$, there are multipliers
$\lambda_{i\J}$ such that
\begin{equation}
\sqJ(S)^\J = \sqJ  \sum_i \lambda_{i\J}(R_{i\J} )^\J \ .
\end{equation}

The equivalent  variational formulation is
\begin{equation} \max_{\{\sqJ\sqdotJ\}}\ \ddt{s(\rho)}
\mbox{ subject to }\ddt{r_i(\rho)}=0 \ {\rm and\ } (\sqJ \sqdotJ |
\sqJ \sqdotJ)_\J=c_\J^2(\rho)\ ,
\end{equation}
where $r_0(\rho)=\Tr\rho$, $r_1(\rho)=\Tr H\rho$ [, $r_2(\rho)=\Tr
N\rho$], and $c_\J^2(\rho)$ are some positive functionals of
$\rho$. The last constraints, one for each subsystem, mean that
each
 subsystem contributes to the overall
 evolution (for the dissipative non-hamiltonian part) by pointing
 towards its \quot{local perception} of the direction of steepest
 (overall) entropy ascent, each with an
 unspecified intensity (which  depends on the values of the
 functionals $c_\J(\rho)$, that are inversely related to the
  internal redistribution characteristic times $\tau_\J(\rho)$).

If two subsystems $A$ and $B$ are non-interacting but in
correlated states, the reduced state operators obey the equations
\begin{eqnarray}
\ddt{\rho_A}&=&-\frac{i}{\hbar}[H_A,\rho_A]+\frac{1}{
\Boltz}\sum_{\stackrel{\scriptstyle J=1}{J\in A}}^r \frac{1}{
2\tau_\J(\rho)} \{ (\Delta M_\J(\rho))^\J,\rho_\J\}\Otimes
(\rho_A)_\Jbar \ , \\
 \ddt{\rho_B}&=&-\frac{i}{\hbar}[H_B,\rho_B] +\frac{1}{
\Boltz}\sum_{\stackrel{\scriptstyle J=1}{J\in B}}^r \frac{1}{
2\tau_\J(\rho)} \{ (\Delta M_\J(\rho))^\J,\rho_\J\}\Otimes
(\rho_B)_\Jbar \ ,
\end{eqnarray}
where $(\rho_A)_\Jbar=\Tr_\J(\rho_A)$,
$(\rho_B)_\Jbar=\Tr_\J(\rho_B)$, and operators $(\Delta
M_\J(\rho))^\J$ result independent of $H_B$  for every $J\in A$
and  independent of $H_A$  for every $J\in B$. Therefore, all
functionals of $\rho_A$ (local observables) remain unaffected by
whatever change in $B$, i.e., locality problems are excluded.

\section{Concluding remarks}

According to QSM and QIT, the uncertainties that are measured by the
physical entropy, are to be regarded as either extrinsic features of
the heterogeneity of an ensemble or as witnesses of correlations with
other systems. Instead, we discuss an alternative theory, QT, based
on  the Hatsopoulos-Gyftopoulos fundamental ansatz \cite{HG,MPLA2}
that also such uncertainties are irreducible (and hence,
\quot{physically real} and \quot{objective} like  standard QM
uncertainties) in that they belong to the state of the individual
system, even  if uncorrelated and even if a member of a homogeneous
ensemble.

According to QT, second law limitations emerge as manifestations of
such additional physical and irreducible uncertainties. The
Hatsopoulos-Gyftopoulos ansatz not only makes a unified theory of QM
and Thermodynamics possible, but gives also a framework for a
resolution of the century old \quot{irreversibility paradox}, as well
as of the conceptual paradox \cite{MPLA2} about the QSM/QIT
interpretation of density operators, which has preoccupied scientists
and philosophers since when Schr\"odinger surfaced it in Ref.
\cite{Schroedinger}. This fundamental ansatz seems to respond to
Schr\"odinger prescient conclusion in Ref. \cite{Schroedinger}:
\quot{\dots in a domain which the present theory (Quantum Mechanics)
does not cover, there is room for new assumptions without necessarily
contradicting the theory in that region where it is backed by
experiment.}

 QT has been described as \quot{an adventurous scheme}
\cite{Nature}, and indeed it requires quite a few conceptual and
interpretational jumps, but (1) it does not contradict any of the
mathematics of either standard QM or TE QSM/QIT, which are both
contained as extreme cases of the unified theory, and (2)  for
nonequilibrium states, no matter how \quot{far} from TE, it offers
the structured, nonlinear equation of motion proposed by this author
which models, deterministically, irreversibility, relaxation and
decoherence, and is based on the additional ansatz of
steepest-entropy-ascent microscopic dynamics.

Many authors, in a variety of contexts \cite{MaxS}, have observed in
recent years that irreversible natural phenomena at all levels of
description seem to obey a principle of general and unifying
validity. It has been named \cite{MaxS} \quot{maximum entropy
production principle}, but we note in this paper that, at least at
the quantum level, the weaker concept of \quot{attraction towards the
direction of steepest entropy ascent} \cite{Frontiers,NATO,thesis} is
sufficient to capture precisely the essence of the second law.

We finally emphasize that the steepest-entropy-ascent, nonlinear law
of motion we propose, and the dynamical group it generates (not just
a semi-group), is a potentially powerful modeling tool that should
find immediate application also outside of QT, namely, regardless of
the dispute about the validity of the Hatsopoulos-Gyftopoulos ansatz
on which QT hinges. Indeed, in view of its well-defined and
well-behaved general mathematical features and solutions, our
equation of motion may be used in phenomenological kinetic and
dynamical theories where there is a need to guarantee full
compatibility with the principle of entropy non-decrease and the
second-law requirement of existence and uniqueness of stable
equilibrium states (for each set of values of the mean energy, of
boundary-condition parameters, and of the mean amount of
constituents).

\end{document}